\documentclass[a4paper,11pt]{article}
\pdfoutput=1 

\usepackage{jheppub} 

\newcommand{\ba}{\begin{eqnarray}}
\newcommand{\ea}{\end{eqnarray}}

\usepackage[T1]{fontenc} 

\title{\boldmath Gravitational AdS to dS phase transition in Einstein-Maxwell-Gauss-Bonnet gravity}

\author[a,b]{Daris Samart}
\author[b,c,d,e,1]{Phongpichit Channuie,\note{Corresponding author.}}

\affiliation[a]{Department of Physics, Faculty of Science, Khon kaen University, 123 Mitraphap road, Khon Kaen, 40002, Thailand}
\affiliation[b]{School of Science, Walailak University, Thasala, Nakhon Si Thammarat, 80160, Thailand}
\affiliation[c]{College of Graduate Studies, Walailak University, Thasala, Nakhon Si Thammarat, \\80160, Thailand}
\affiliation[d]{Research Group in Applied, Computational and Theoretical Science (ACTS), Walailak University, Thasala, Nakhon Si Thammarat, 80160, Thailand}
\affiliation[e]{Thailand Center of Excellence in Physics, Ministry of Education, Bangkok 10400, Thailand}

\emailAdd{darisa@kku.ac.th, channuie@gmail.com}

\abstract{In this work, we revisit a toy model proposed by Camanho {\it et.al.} [JHEP 10 (2015) 179] and extensively study the possible existence of gravitational phase transition from AdS to dS geometries by adding the Maxwell field as an impurity substitution. We show that the phase transitions proceed via the bubble nucleation of spherical thin-shells described by different branches of the solution which host a black hole in the interior. In order to demonstrate the existence of the phase transition, we examine how the free energy and temperature depend on the higher-order gravity coupling $(\lambda)$ indicating the possibility of thermalon mediated phase transition. We observe that the phase transitions of the charged case is possible in which the required (maximum) temperature is lower than that of the neutral case. Interestingly, we also discover that the critical temperature and the coupling $\lambda$ of the phase transitions are modified when having the charge. Notably, our results agree with the claim that the generalized gravitational phase transition is a generic behavior of the higher-order gravity theories even the matter field is added.}

\begin{document} 
\maketitle
\flushbottom

\section{Introduction}

One of the greatest challenges in physics nowadays is to explain the positive value of the cosmological constant or, equivalently, the energy density of the vacuum. Regarding the positiveness of the cosmological constant, phase transition in gravitational physics posses one of the interesting subjects for decades. Indeed, phase transitions between two competing vacua of a given theory are quite common in physics. They occur when the free energy of the actual vacuum becomes greater than the other due to a variation of some parameter of the system. Phase transitions between two competing vacua with different cosmological constants have been so far discussed in the context of gravitational instantons \cite{Coleman1977,Coleman1980}. In various proposed gauge/gravity dualities, another important example of the gravitational phase transitions is known as the Hawking-Page transition \cite{Hawking:1982dh}. This is the first-order phase transition competing between thermal AdS space and the Schwarzschild-AdS black hole. In addition, a number of publications of phase transition in context of the AdS and dS black hole thermodynamics have been actively studied in several aspects and various models of the higher-order gravity \cite{Nojiri:2002qn,Nojiri:2001aj,Nojiri:2017kex,Wiltshire:1985us,Chamblin:1999tk,Cai:2001dz,Cai:2003gr,Charmousis:2008kc,Chernicoff:2016jsu,Castro:2013pqa,Mishra:2019ged,Guo:2018exx,Camanho:2011rj,Cai:2013qga,EslamPanah:2017yoc,Wei:2012ui,Chakraborty:2015taq,Cvetic:2003zy,Garraffo:2008hu,Camanho:2013kfa}

More interestingly in higher-order theories of gravity, a number of recent studies have focused on thermalon mediated phase transitions \cite{Camanho:2012da,Camanho:2015zqa,Camanho:2013uda} in many cases of Lovelock gravity with a vacuum solution. These types of phase transitions proceed through the nucleation of the spherical thin-shell bubbles, so-called thermalon which is the Euclidean sector of a static bubble. This thin-shell stays between two regions described by different branches of the solution which host the black hole in the interior. On the other hand, the thermalon is a finite temperature instanton which is considered as a thermodynamic phase and described an intermediate state. In a finite time, when the thermalon forms, it is dynamically unstable and then expands to occupy entire space. Hence this effectively changes the asymtotic structure of the spacetime. Once the cosmological constant is fixed, it was shown in Refs.\cite{Camanho:2015zqa,Nojiri:2001pm,Cvetic:2001bk} that thermal AdS space underwent a thermalon-mediated phase transition to an asymptotically dS black hole geometry.\\  
\indent It has been found that this type of the gravitational phase transition is a generalized phase transition of the Hawking-Page mechanism in Lovelock gravity. In addition, the AdS to dS gravitational phase transition is claimed to be a generic behavior of the higher order of the gravitational theories \cite{Camanho:2015zqa}. However, the inclusion of the matter in this toy model of the phase transition has not been studied yet. It is worth to investigate the phase transition profile of the model with the matter field.\\
Moreover, the Lovelock or higher order gravity naturally arises in string theory. Therefore, a study of the phase transitions in this type of gravity might reveal some interesting features of the consequences in the string theory at low energy regimes. In particular, it is expected that we can gain a better understanding of the phase transition phenomenon in the AdS/CFT correspondence paradigms. Although the AdS/CFT is extensively studied in various aspects and its nature is widely known, the dS/CFT counterpart is less studied and poorly understood. For this reason, the study of the AdS to dS phase transition in this work may be also useful for uncovering the nature of the dS/CFT. In the present work, we therefore revisit a toy model proposed by Ref.\cite{Camanho:2015zqa} and extensively study the possible existence of gravitational phase transition from AdS to dS geometries by adding the Maxwell field. In addition, we also investigate the effects of the static charge on the critical temperature and the coupling of the higher-order gravity term, the Gauss-Bonnet term in this work, causing the phase transition.\\
\indent The content of the paper is organized as follows. In section \ref{s2}, we will review some basics of Lovelock gravity in the vierbein formalism \cite{Camanho:2015zqa} with the Maxwell field that are the starting point for the computations of the present work and construct a junction condition of Lovelock-Maxwell gravity. We then focus on a special case of Lovelock gravity in which the action is reduced to Einstein-Gauss-Bonnet-Maxwell (EGBM) gravity. In this section, we also derive the effective potential of the thermalon EMGB gravity and examine the thermalon solutions as well as the stability and dynamics of the thermalon. In section \ref{s3}, we study the gravitational phase transition and the relevant thermodynamic quantities in EMGB gravity. Here we examine how the free energy and temperature depend on the coupling indicating the possibility of thermalon mediated phase transition. We conclude our findings in the last section.
\section{Formalism}
\label{s2}
\subsection{Lovelock-Maxwell gravity action}
We start with the action of the Lovelock gravity in vierbein formalism with inclusion of the Maxwell field in $d$ dimensions, it reads \cite{Camanho:2013kfa,Camanho:2015ysa},
\allowdisplaybreaks
\begin{eqnarray}
\mathcal{I} = \frac{1}{16\pi G_N(d-3)!}\left[\, \sum_{k=0}^K\,\frac{c_k}{d-2\,k}\left( \int_{\mathbb{M}} \mathcal{L}_k - \int_{\mathbb{\partial M}}\mathcal{B}_k \right) + \int_{\mathbb{M}} \mathcal{F}\wedge *\mathcal{F} \right],
\end{eqnarray}
where $\mathbb{M}$ and $\partial\mathbb{M}$ are the spacetime manifold and its boundary, respectively. In this work, all ingredients of the Lovelock gravity in vierbein formalisms are given by,
\allowdisplaybreaks
\begin{eqnarray}
\mathcal{L}_k &=& \epsilon_{a_1\cdots a_d}\,R^{a_1 a_2}\wedge \cdots \wedge R^{a_{2k-1}a_{2k}}\wedge e^{a_{2k+1}}\wedge\cdots\wedge e^{a_d}\,,
\\
\mathcal{B}_k &=& k\int_0^1 d\xi \,\epsilon_{a_1\cdots a_d}\,\theta^{a_1 a_2}\wedge \mathfrak{F}^{a_3 a_4}\wedge \cdots \wedge \mathfrak{F}^{a_{2k-1}a_{2k}}\wedge e^{a_{2k+1}}\wedge\cdots\wedge e^{a_d}\,,
\\
R^{ab} &=& d\omega^{ab} + \omega^a_{~\,c}\wedge \omega^{cb}\,,
\\
\mathfrak{F}^{ab} &=& R^{ab} + (\xi^2-1)\,\theta^a_{~\,c}\wedge \theta^{cb}\,,
\\
\theta^{ab} &=& \left( n^a\,K^b_{~\,c} - n^b\,K^a_{~\,c}\right)e^c\,,
\end{eqnarray}
where $R^{ab}$ is the curvature two-form with $\omega^{ab}$, the torsionless Levi-Civita spin connection. Moreover, $e^a = e^a_{~\,\mu}dx^\mu$\,, $n^a$ and $K^{ab}$ are the vierbein one-form, normal unit vector and extrinsic curvature, respectively. The spherically symmetric solution of the theory is taken form,
\begin{eqnarray}
ds^2 = -\,f(r)\,dt^2 + \frac{dr^2}{f(r)} + r^2\,d\Omega^2_{(\sigma)\,d-2}\,,
\end{eqnarray}
where $d\Omega^2_{(\sigma)\,d-2}$ is the line element of the $(d-2)$-dimensional surface of the constant curvature, $\sigma$ with $\sigma = 1,~0,~-1$ (spherical, flat and hyperbolic geometries, respectively). More importantly, we will use the normalization of the gravitational constant such that $16\pi G_N(d-3)!=1$ \cite{Camanho:2015zqa,Camanho:2013uda}. The equation of motion of the Maxwell field in the vacuum is given by
\begin{eqnarray}
d*\mathcal{F} = 0\,\qquad {\rm and} \qquad d\mathcal{F} = 0\,,\qquad {\rm with}\qquad \mathcal{F} = dA\,,
\end{eqnarray}
where $A$ is the vector potential one-form. The field strength tensor $\mathcal{F}$ is given by the following ansatz,
\begin{eqnarray}
\mathcal{F} = \frac{Q}{r^{d-1}}\,dt\wedge dr\,,
\end{eqnarray}
where $Q$ is the electric charge. Having use all ingredients introduced, we can write the solution of the theory by introducing the following polynomial as
\begin{eqnarray}
\Upsilon[g] &=& \sum_{k=0}^K c_k\,g^k = \frac{\mathcal{M}}{r^{d-1}} - \frac{\mathcal{Q}^2}{r^{2(d-2)}}\,,
\\
g &\equiv& g(r) = \frac{\sigma - f(r)}{r^2}\,.
\end{eqnarray}
The parameters $\mathcal{M}$ and $\mathcal{Q}$ are related to the black hole ADM mass ($M$) and the electric charge ($Q$) via,
\begin{eqnarray}
\mathcal{M} = \frac{\Gamma\big(\frac{d}{2}\big)\,M}{(d-2)!\,\pi^{\frac{d}{2}-1}}\,,\qquad \mathcal{Q}^2 = \frac{Q^2}{(d-2)(d-3)}\,.
\end{eqnarray}
We refer the detail derivation of the $\Upsilon$ solution in Refs. \cite{Charmousis:2008kc,Garraffo:2008hu,Castro:2013pqa,Chernicoff:2016jsu}.

\subsection{Junction condition in Lovelock-Maxwell gravity}
In this work, the main purpose it to study the dynamics of unstable spherical thin shell (thermalon) of the Lovelock-Maxwell gravity. To do this, we firstly divide the manifold of the spacetime into two regions. We focus in the case of timelike surface of the manifold. Then the manifold is decomposed as $\mathbb{M} = \mathbb{M}_{-} \cup (\Sigma\times\xi) \cup \mathbb{M}_+$ where $\Sigma$ is the junction hypersurface of two regions of the manifolds and $\xi \in [0,1]$ is interpolating both regions. The $\mathbb{M}_+$ and $\mathbb{M}_-$ are outer and inner regions of the manifolds, respectively. The metric tensor, $f_\pm(r)$ are also used to describe geometries of the outer and inner manifolds. One writes two different line elements of the spacetimes that is used to describe AdS outer ($+$) and dS inner ($-$) spacetime as
\begin{eqnarray}
ds_\pm^2 = -f_\pm(r_\pm)\,dt_\pm^2 + \frac{dr_\pm^2}{f_\pm(r_\pm)} + r^2_\pm\,d\Omega_{(\sigma),\,d-2}^2\,,
\label{out-in-line}
\end{eqnarray}
again $\pm$ correspond to outer and inner spacetimes respectively. In the latter, we will focus our study in 5-dimensional spacetime. This gives $d\Omega_{(\sigma),\,d-2}^2 \stackrel{d=5}{=} d\Omega_{(\sigma),\,3}^2$ and it is defined by,
\begin{eqnarray}
d\Omega_{(\sigma),\,d-2}^2 =
\begin{cases}
d\theta^2 + \sin^2\theta\, d\chi^2 + \sin^2\theta\,\sin^2\chi\,d\phi^2 : \sigma =1\,,
\\
d\theta^2 + d\chi^2 + d\phi^2 : \sigma =0\,,
\\
d\theta^2 + \sinh^2\theta\, d\chi^2 + \sinh^2\theta\,\sinh^2\chi\,d\phi^2 : \sigma =-1\,,
\end{cases}
\end{eqnarray}
Next step, we construct a manifold $\mathbb{M}$ by matching $\mathbb{M}_\pm$ at their boundaries. We choose
the boundary hypersurfaces $\partial \mathbb{M}_\pm$ as
\begin{eqnarray}
\partial \mathbb{M}_\pm := \Big\{ r_\pm = a | f_\pm > 0\,\Big\}
\end{eqnarray}
with parameterizations of the coordinates
\begin{eqnarray}
r_\pm = a(\tau)\,,\qquad\qquad t_\pm = \widetilde{t}_\pm(\tau)\,,
\end{eqnarray}
where $\tau$ is comoving time of the induced line elements of the hypersurface ($\Sigma$) which takes the same form in both of two manifolds $\mathbb{M}_\pm$ at the boundaries, it reads,
\begin{eqnarray}
ds_\Sigma^2 = -d\tau^2 + a^2(\tau)\,d\Omega_{(\sigma),\,d-2}^2\,.
\label{surface-line}
\end{eqnarray}
Applying the coordinate parameterizations to the line elements of the manifolds in Eq. (\ref{out-in-line}), one finds,
\begin{eqnarray}
ds_\pm^2 &=& -f_\pm(a)\,d\,\widetilde{t}_\pm(\tau)^2 + \frac{da(\tau)^2}{f_\pm(a)} + r^2_\pm\,d\Omega_{(\sigma),\,d-2}^2\,,
\nonumber\\
&=& -\left(f_\pm(a)\left(\frac{\partial\, \widetilde{t}_\pm}{\partial \tau}\right)^2 - \frac{1}{f_\pm(a)}\left(\frac{\partial a}{\partial \tau}\right)^2 \right) d\tau^2 + a^2(\tau)\,d\Omega_{(\sigma),\,d-2}^2\,.
\label{rewrite-out-in-line}
\end{eqnarray}
As mentioned earlier, the line elements of the hypersurface must has the same form at the boundaries of the manifolds. Then we compare line elements in Eqs. (\ref{surface-line}) and (\ref{rewrite-out-in-line}), we obtain the following constraint,
\begin{eqnarray}
1 = f_\pm(a)\left(\frac{\partial\, \widetilde{t}_\pm}{\partial \tau}\right)^2 - \frac{1}{f_\pm(a)}\left(\frac{\partial a}{\partial \tau}\right)^2\,.
\end{eqnarray}

It has been shown in detail in Refs \cite{Camanho:2013uda,Camanho:2015ysa} that the continuity of the junction condition across the hypersurface in electro-the vacuum case is written in terms of the canonical momenta, $\pi_{AB}^\pm$ as
\begin{eqnarray}
\pi_{AB}^+ - \pi_{AB}^- = 0\,.
\label{junction-eqn}
\end{eqnarray}
We note that the capital Latin alphabets $A,\,B,\,C,\,\cdots$ are the veirbein indices of the hypersurface, $\Sigma$\,.
The canonical momentum, $\pi_{AB}$ is derived by varying the gravitational action of the boundary with respect to the induced metric, $h_{ab}$ on the hypersurface, $\Sigma$ i.e. \cite{Camanho:2013uda,Camanho:2015ysa},
\begin{eqnarray}
\delta \mathcal{I}_{\partial\mathbb{M}} = -\int_{\partial\mathbb{M}}d^{(d-1)}x\,\pi_{AB}\,\delta h^{AB}\,,
\end{eqnarray}
where the canonical momentum, $\pi^A_{~\,B}$ is given by
\begin{eqnarray}
\pi^A_{~\,B} &=& -\,\frac12\,\frac{\delta \mathcal{I}_{\partial\mathbb{M}}}{\delta\,e^B}\,\wedge\,e^A
\nonumber\\
&=& \sum_{k=1}^K\,k\,c_k \int_0^1 d\xi\,K^{A_1}\,\wedge\,\mathfrak{F}^{A_2\,A_3}(\xi)\,\wedge\,\cdots\mathfrak{F}^{A_{2k-2}\,A_{2k-1}}(\xi)
\,e^{A_{2k}\,\cdots\,A_{d-2}\,A}\,\epsilon_{A_1\,\cdots\,A_{d-2}\,B}\,,
\end{eqnarray}
with $K^A = K^A_{~\,B}\,e^B$\,.

It has been also demonstrated in Refs. \cite{Camanho:2013uda,Camanho:2015ysa} that for our study case in the latter, the diagonal components of the $\pi_{ab}^\pm$ give the some relation between time and spatial parts via the following constraint,
\begin{eqnarray}
\frac{d}{d\,\tau}\left( a^3\,\pi_{\tau\tau}^\pm\right) = 3\,a^2\,\dot a\,\pi_{\varphi_i\varphi_i}^\pm\,,\qquad \varphi_i = \varphi_1\,,\,\varphi_2\,,\,\varphi_3 = \theta\,,\,\chi\,,\,\phi\,.
\end{eqnarray}
In addition, the (co-moving) time component of the $\pi_{ab}^\pm$ is re-written in the compact form as \cite{Camanho:2015zqa,Camanho:2013uda,Camanho:2015ysa},
\begin{eqnarray}
\Pi^{\pm} = \pi_{\tau\tau}^\pm = \frac{\sqrt{\dot a^2 + f_\pm (r)}}{a}\int_0^1d\xi\,\Upsilon'\Bigg[ \frac{\sigma - \xi^2 f_\pm(a) + (1-\xi^2)\,\dot a^2}{a^2} \Bigg]\,,
\label{bigpi-solution}
\end{eqnarray}
where $\Upsilon'[x] = d\Upsilon[x]/dx$\,. Furthermore, it is convenient to define new variables, $\widetilde{\Pi} = \Pi^+ - \Pi^-$, the junction conditions of the continuity across hypersurface are given by,
\begin{eqnarray}
\widetilde{\Pi} = 0 = \frac{d\widetilde{\Pi}}{d\tau}\,.
\end{eqnarray}
Refs. \cite{Camanho:2013uda,Camanho:2015ysa} have derived the further compact form of $\widetilde{\Pi}$ as
\begin{eqnarray}
\widetilde{\Pi} = \int_{\sqrt{H-g_-}}^{\sqrt{H-g_+}} dx\,\Upsilon'\big[ H- x^2\big]\,,
\end{eqnarray}
where $H = (\sigma + \dot a^2)/a^2$. We will specific the Lovelock-Maxwell gravity at $K=2$ and this leads to a so-called Einstein-Gauss-Bonnet-Maxwell (EGBM) gravity in the latter. From now on, we will work on the Euclidean signature, i.e. $t\,\to\,i\,t$ for studying the thermalon which is the Euclidean sector of the spherical bubble thin-shell. This gives $\dot a^2 \,\to\,-\dot a^2$ and $\ddot a \,\to\,-\ddot a$\,.

\subsection{The Einstein-Gauss-Bonnet-Maxwell gravity}
The EGBM gravity is a reduction form of the Lovelock-Maxwell gravity at $K=2$, and one finds
\begin{eqnarray}
\mathcal{I} &=&  \sum_{k=0}^{K=2}\,\frac{c_k}{d-2\,k}\left( \int_{\mathbb{M}} \mathcal{L}_k - \int_{\mathbb{\partial M}}\mathcal{B}_k \right) + \int_{\mathbb{M}} \mathcal{F}\wedge *\mathcal{F}
\nonumber\\
&=& \int d^dx\left[ -\,\varepsilon_\Lambda\,\frac{(d-1)(d-2)}{L^2} + R + \frac{\lambda\,L^2}{(d-3)(d-4)}\Big( R^2 - 4\,R_{ab}\,R^{ab} + R_{abcd}\,R^{abcd} \Big) \right] 
\nonumber\\
&& -\, \frac{1}{4}\int d^dx\,\mathcal{F}_{ab}\,\mathcal{F}^{ab}
\nonumber\\
&& -\, \int_{\mathbb{\partial M}} d^{(d-1)}x\sqrt{-h}\left[ K +\frac{2\,\lambda\,L^2}{(d-3)(d-4)}
\left\{ J - 2\left( \mathcal{R}^{AB} -\frac12\,h^{AB}\,\mathcal{R}\right)K_{AB}\right\}\right] \,,
\label{EGBM-action}
\end{eqnarray}
where $J\equiv h^{AB}\,J_{AB}$ is the trace of $J_{AB}$ which is builded up from $K_{AB}$ as
\begin{eqnarray}
J_{AB} = \frac13\left( 2\,K\,K_{AC}\,K_B^C + K_{CD}\,K^{CD}\,K_{AB} - 2\,K_{AC}\,K^{CD}\,K_{DB} - K^2\,K_{AB}\right) ,
\end{eqnarray}
and $\mathcal{R}_{AB}$ is the Ricci tensor (intrinsic curvature) of the hypersurface, $\Sigma$\,. More importantly, we note that the coefficients of the Lovelock theory for the Gauss-Bonnet gravity case are given by
\begin{eqnarray}
c_0 = \frac{1}{L^2}\,,\qquad c_1 = 1\,,\qquad c_2 = \lambda\,L^2\,.
\end{eqnarray}
Since we have identified the cosmological constant $(\Lambda)$ and normalized $16\,\pi\,(d-3)\,\,G$ the parameters of the theory as
\begin{eqnarray}
\Lambda = \varepsilon_\Lambda\,\frac{(d-1)\,(d-2)}{2\,L^2}\,,\qquad 16\,\pi\,G\,(d-3) ! = 1 \,,
\end{eqnarray}
where $\varepsilon_\Lambda = \pm\,1$\, is the sign of the bare cosmological constant and we use the $\varepsilon_\Lambda = +\,1$ (de-Sitter) of the bare cosmological constant in this work.
The solution of the polynomial, $\Upsilon[g]$ in the EMGB theory is given by
\begin{eqnarray}
\Upsilon[g] = -\frac{1}{L^2} + g + \lambda\,L^2\,g^2 = \frac{\mathcal{M}}{r^{d-1}} - \frac{\mathcal{Q}^2}{r^{2(d-2)}}\,.
\end{eqnarray}
One finds the solutions of $g$ from the above equation as
\begin{eqnarray}
g_\pm \equiv g_\pm(r) = -\,\frac{1}{2\,\lambda\,L^2}\left( 1 \pm \sqrt{1+ 4\,\lambda\left[ 1
+ L^2\left( \frac{\mathcal{M}}{r^{d-1}} - \frac{\mathcal{Q}^2}{r^{2(d-2)}} \right)\right]}\,\right).
\end{eqnarray}
Therefore, the solutions of the line elements for inner and outer manifolds in Eq. (\ref{rewrite-out-in-line}) are given by \cite{Wiltshire:1985us,Charmousis:2008kc}
\begin{eqnarray}
f_\pm \equiv f_\pm(r) = \sigma + \frac{r^2}{2\,\lambda\,L^2}\left( 1 \pm \sqrt{1+ 4\,\lambda\left[ 1
+ L^2\left( \frac{\mathcal{M}}{r^{d-1}} - \frac{\mathcal{Q}^2}{r^{2(d-2)}} \right)\right]}\,\right).
\label{metric-f-pm}
\end{eqnarray}

Next we turn to construct the junction condition of EMGB gravity. One recalls the compact form of the co-moving time component of the canonical momenta, given by $\Pi = \pi_{\tau\tau}$ in Eq. (\ref{bigpi-solution}) as \cite{Camanho:2015zqa,Camanho:2013uda,Camanho:2015ysa}
\begin{eqnarray}
\Pi_\pm &=& \frac{\sqrt{\dot a^2 + f_\pm (a)}}{a}\int_0^1 d\xi\,\Upsilon'\Bigg[ \frac{\sigma - \xi^2 f_\pm(a) + (1-\xi^2)\,\dot a^2}{a^2} \Bigg]
\nonumber\\
&=& \frac{2}{3}\,\lambda\,L^2\,g_\pm \sqrt{H-g_\pm} + \frac{4}{3}\,H\,\lambda\,L^2\sqrt{H-g_\pm} + \sqrt{H-g_\pm}\,,
\label{integrate-pi}
\end{eqnarray}
where the following definitions have been used to perform the above integration,
\begin{eqnarray}
H(a,\,\dot a) &=& \frac{\sigma + \dot a^2}{a^2}\,,
\\
g(a) &=& \frac{\sigma - f(a)}{a^2}\,,
\\
\Upsilon'[x] &=& \frac{d\,\Upsilon[x]}{d\,x} 
= 1 + 2\,\lambda\,L^2\,x \,.
\end{eqnarray}
The junction condition of the EMGB gravity in Eq. (\ref{junction-eqn}) is implied that
\begin{eqnarray}
\widetilde{\Pi} = \Pi_+ - \Pi_- = 0\,\quad \Longrightarrow \,\quad \Pi_+^2 = \Pi_-^2\,.
\label{junction-EMGB}
\end{eqnarray}
Substituting the results of the $\Pi_\pm$ in Eq. (\ref{integrate-pi}) in the junction condition above, we find
\allowdisplaybreaks
\begin{eqnarray}
-\dot a^2 &=& \frac{a^{d+1} }{12 \lambda\,  L^2 }\,\frac{\left(g_+ \left(2 g_+ \lambda\,  L^2+3\right)^2-g_- \left(2 g_- \lambda  L^2+3\right)^2\right)}{(\mathcal{M}_+ -\mathcal{M}_-)-\left(\mathcal{Q}_+^2-\mathcal{Q}_-^2\right)/a^{d-3} } +\sigma \,,
\end{eqnarray}
where we have used the following identity in the last line for the denominator,
\begin{eqnarray}
g_\pm + \lambda\,L^2\,g_\pm^2 = \frac{1}{L^2} + \frac{\mathcal{M}_\pm}{a^{d-1}} - \frac{\mathcal{Q}_\pm^2}{a^{2(d-2)}}\,.
\end{eqnarray}
The junction condition equation may be rewritten in terms of kinetic and effective potential energies as
\begin{eqnarray}
\Pi_+^2 = \Pi_-^2\,\quad \Longleftrightarrow \,\quad\, \frac12\,\dot a^2 + V(a) = 0\,.
\end{eqnarray}
Then the effective potential $V(a)$ of the junction condition equation is given by
\begin{eqnarray}
V(a) &=& \frac{a^{d+1} }{24 \lambda\,  L^2 }\,\frac{\left(g_+ \left(2 g_+ \lambda\,  L^2+3\right)^2-g_- \left(2 g_- \lambda  L^2+3\right)^2\right)}{(\mathcal{M}_+ -\mathcal{M}_-)-\left(\mathcal{Q}_+^2-\mathcal{Q}_-^2\right)/a^{d-3} } + \frac{\sigma}{2} \,,
\label{V-form1}
\end{eqnarray}
Moreover, one always can reduce the power of the $g_\pm$ functions via the following identities,
\begin{eqnarray}
g_\pm^3 = \frac{g_\pm }{\lambda\,  L^2}\left(\frac{\mathcal{M}_\pm}{a^{d-1}}-\frac{\mathcal{Q}_\pm^2}{a^{2 (d-2)}}-g_\pm+\frac{1}{L^2} \right),\;
g_\pm^2 = \frac{1}{\lambda \, L^2}\left(\frac{\mathcal{M}_\pm}{a^{d-1}}-\frac{\mathcal{Q}_\pm^2}{a^{2 (d-2)}}- g_\pm + \frac{1}{L^2}\right) .
\end{eqnarray}
Using the power reductions of the $g_\pm$\,, one re-write the effective potential $V(a)$ in Eq. (\ref{V-form1}) at the first order of $g_\pm$ as,
\begin{eqnarray}
V(a) &=& \frac{a^{d+1}}{ 24\, \lambda\,  L^2 \Big[(\mathcal{M}_+ -\mathcal{M}_-)-\left(\mathcal{Q}_+^2-\mathcal{Q}_-^2\right)/a^{d-3} \Big] }
\nonumber\\
&&\qquad\times\,\Bigg[ \left(1+ 4\,\lambda\right) g + 4 \left(2 + g \lambda\,  L^2\right)\left(\frac{\mathcal{M}}{a^{d-1}} - \frac{\mathcal{Q}^2}{a^{2(d-2)}}\right)\Bigg]\Bigg|_-^+ \,,
\label{V-form2}
\end{eqnarray}
where the symbol $\big[ \mathcal{O}\big]\big|_-^+$ is defined by
\begin{eqnarray}
\big[ \mathcal{O}\big]\big|_-^+ \equiv \mathcal{O}_+ - \mathcal{O}_-\,.
\end{eqnarray}

\begin{figure}[h]	
	\includegraphics[width=15cm]{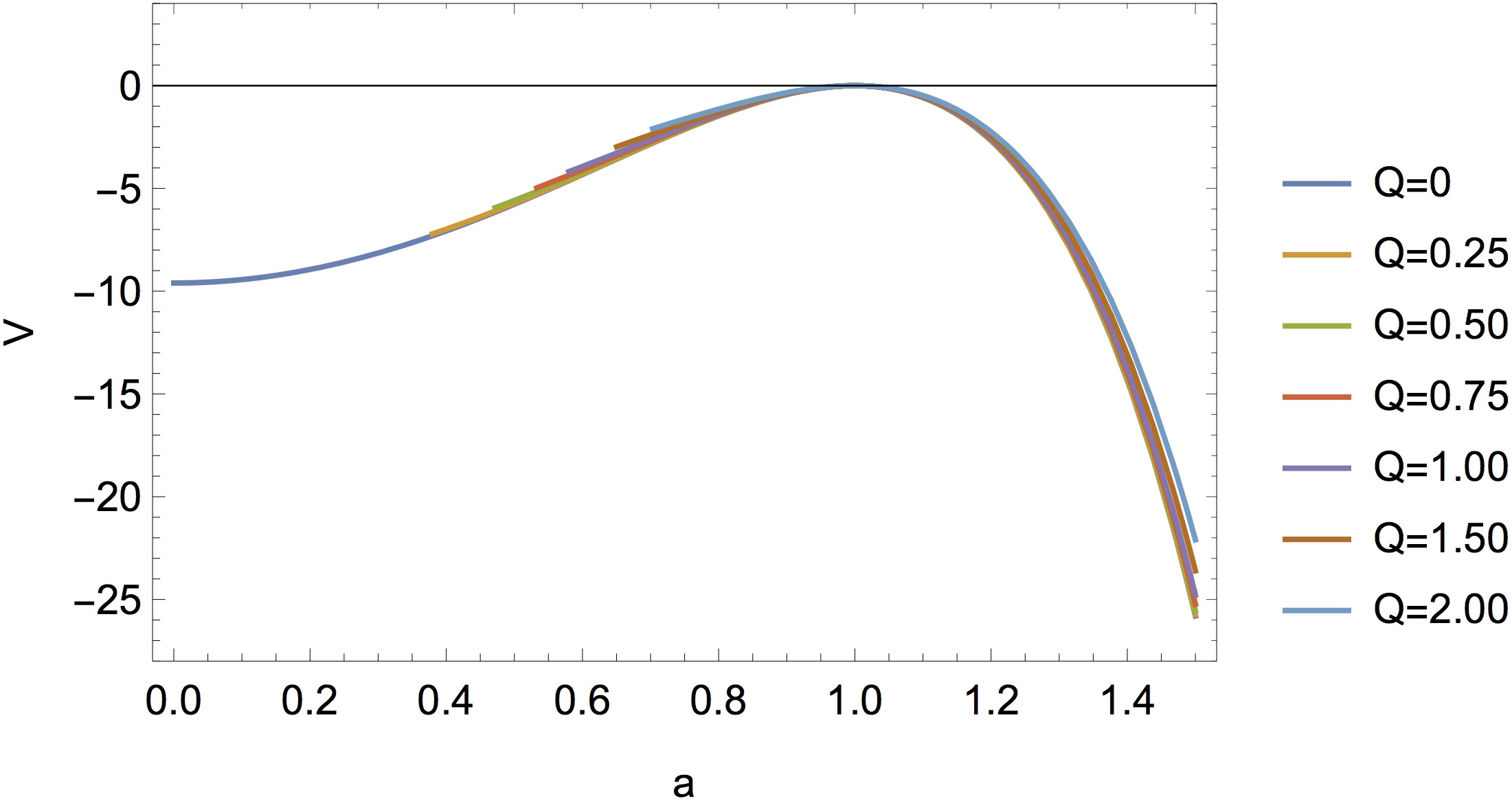}
	\centering
	\caption{The figure displays the shapes of the effective potential of thermalon in various values of the charge $\mathcal{Q}$ with $\lambda = 0.015$, $a_\star=1$, $L=1$, $d=5$ and $\sigma=1$.}
	\label{thermalon-potential}
\end{figure}
We continue to evaluate the derivative of the effective potential, $V'(a)$ and it reads,
\begin{eqnarray}
V'(a) &=& \frac{a^d}{24\, \lambda\,  L^2 \left(\mathcal{M}_+ -\mathcal{M}_- - \left(\mathcal{Q}_+^2-\mathcal{Q}_-^2\right)/a^{d-3} \right)}
\nonumber\\
&&\quad\times\,\Bigg[(1+d)\,(1+ 4\,\lambda)\,g -\left( d-17 +2( d -5)\lambda\, L^2 g\right)\frac{\mathcal{M}}{a^{d-1}} 
\nonumber\\
&&\qquad\qquad\qquad\qquad\qquad\; +\, 2 \left(5\,d - 22 + 2(2\, d - 7)\,\lambda\, L^2 g \right)\frac{\mathcal{Q}^2}{a^{2(d-2)}} \Bigg]\Bigg|_-^+
\nonumber\\
&+& \frac{a^3 (3-d) \left(\mathcal{Q}_+^2-\mathcal{Q}_-^2\right) }{24\, \lambda\,  L^2 \left(\mathcal{M}_+ -\mathcal{M}_- - \left(\mathcal{Q}_+^2-\mathcal{Q}_-^2\right)/a^{d-3} \right)^2}
\nonumber\\
&&\quad\times\left[( 1+ 4\,\lambda)\,g + 4 \left(2 + \lambda\,L^2 g \right)\left(\frac{\mathcal{M}}{a^{d-1}}
- \frac{\mathcal{Q}^2}{a^{2(d-2)}} \right)\right]\Bigg|_-^+\,.
\label{div-eff-potential}
\end{eqnarray}
To eliminate $g'$, we have used following identities,
\begin{eqnarray}
g' &=& \frac{1}{\Upsilon'[g]}\left( (1-d)\,\frac{\mathcal{M}}{a^d} - 2\,(2-d)\,\frac{\mathcal{Q}^2}{a^{2d-3}} \right),
\\
\Upsilon'[g] &=& 1 + 2\,\lambda\,L^2\,g\,.
\end{eqnarray}

We note that if we drop the static charge, i.e. $\mathcal{Q}_\pm = 0$\,. The effective potential and its derivative are reduced to the same forms as the neutral case that given in Refs. \cite{Camanho:2013uda,Hennigar:2015mco} i.e.,
\begin{eqnarray}
V(a) &\stackrel{\mathcal{Q}_\pm\to 0}{=}& \frac{a^{d+1}}{ 24\, \lambda\,  L^2 \Big[(\mathcal{M}_+ -\mathcal{M}_-) \Big] }\,\Bigg[ \left(1+ 4\,\lambda\right) g + 4 \left(2 + g \lambda\,  L^2\right)\frac{\mathcal{M}}{a^{d-1}} \Bigg]\Bigg|_-^+ + \frac{\sigma}{2}
\\
V'(a) &\stackrel{\mathcal{Q}_\pm\to 0}{=}& \frac{a^d}{24\, \lambda\,  L^2 \left(\mathcal{M}_+ -\mathcal{M}_- \right)}
\left[(1+d)\,(1+ 4\,\lambda)\,g -\left( d-17 +2( d -5)\lambda\, L^2 g\right)\frac{\mathcal{M}}{a^{d-1}}  \right]\Bigg|_-^+ \,.
\nonumber\\
\end{eqnarray}

According to Ref. \cite{Giribet:2019dmg}, it has been shown that the continuity of the (electric) static charge across the hypersurface in the EMGB gravity gives
\begin{eqnarray}
\mathcal{Q} = \mathcal{Q}_+ = -\,\mathcal{Q}_-\,.
\end{eqnarray}
Having use above continuity equation, the effective potential of the thermalon dynamics and its derivative become
\begin{eqnarray}
V(a) &=& \frac{a^{d+1}}{ 24\, \lambda\,  L^2 \big(\mathcal{M}_+ -\mathcal{M}_-\big) }\,\Bigg[ \left(1+ 4\,\lambda\right) g
+ 4 \left(2 + \lambda\,  L^2 g \right)\left(\frac{\mathcal{M}}{a^{d-1}} - \frac{\mathcal{Q}^2}{a^{2(d-2)}}\right)\Bigg]\Bigg|_-^+ + \frac{\sigma}{2}\,,
\nonumber\\
\label{final-charge-eff-potential}
\\
V'(a) &=& \frac{a^d}{24\, \lambda\,  L^2 \big(\mathcal{M}_+ -\mathcal{M}_- \big)}
\nonumber\\
&&\quad\times\,\Bigg[(1+d)\,(1+ 4\,\lambda)\,g -\left( d-17 +2( d -5)\lambda\, L^2 g\right)\frac{\mathcal{M}}{a^{d-1}} 
\nonumber\\
&&\qquad\qquad\qquad\qquad\qquad\qquad\qquad +\, 4\,(2\, d - 7)\,\lambda\, L^2 g \,\frac{\mathcal{Q}^2}{a^{2(d-2)}} \Bigg]\Bigg|_-^+\,.
\label{final-charge-div-eff-potential}
\end{eqnarray}
We close this subsection by discussing the effect of the Maxwell field (static charge) on the thermalon effective potential displaying in figure \ref{thermalon-potential}. One sees that the inclusion of the static charge does not change the shape of the effective potential except the existences of of the potential. Increasing of the static charge makes the existences of the potential closer to the thermalon position as shown in figure \ref{thermalon-potential}. In addition, setting $\mathcal{Q}=0$, we precisely reproduce the effective potential of the thermalon in the neutral case as done in Ref.\cite{Camanho:2013uda}.

\subsection{Thermalon dynamics of the EMGB gravity and its stability}
We have derived the effective potential of the thermalon and its derivative in the previous section for the EMGB gravity. Now we are going to work out the thermalon solutions as well as investigating the stability and dynamics of the thermalon. We firstly consider the solutions of the thermalon configuration by imposing $V(a_\star) = 0 = V'(a_\star)$\,. Solving those two equations, one obtains the solutions of $\mathcal{M}_\pm$ in terms of $g_\pm^\star$, $a_\star$, $\lambda$, $L$, $d$ and $\mathcal{Q}$ as,
\begin{eqnarray}
\mathcal{M}_+(g_\pm^\star,\,a_\star,\,\lambda,\,L^2,\,\mathcal{Q}^2)\; \equiv\, \mathcal{M}_+^\star 
&=& \frac{1}{4\, \lambda\,L^2\left(a_\star^2 (d-1)+2 (d-5) \lambda\,L^2 \sigma \right)}
\nonumber\\
&&\times\Big[a_\star^{d+1}\, (d-1)\,(1 + 4\,\lambda)  \left(3 + 2\,\lambda\,L^2\,g_-^\star \right) 
\nonumber\\
&&\quad\; +\, 4 \,a_\star^{d-1}\,(d+1)\,(1 + 4\, \lambda )\, \lambda\,L^2\, \sigma
\nonumber\\
&&\quad\; +\, 4\, a_\star^{5-d} \left(5\,d-13 + 2 (d-3)\,\lambda\,L^2\, g_-^\star\right)\lambda\,L^2\, \mathcal{Q}^2 
\nonumber\\
&&\quad\; +\, 16\, a_\star^{3-d}\, (2 d-7)\, \lambda ^2\,L^4\, \mathcal{Q}^2 \sigma \Big],
\label{M_+}\\
\mathcal{M}_-(g_\pm^\star,\,a_\star,\,\lambda,\,L^2,\,\mathcal{Q}^2)\; \equiv\, \mathcal{M}_-^\star &=& \frac{1}{4\, \lambda\,L^2\left(a_\star^2 (d-1)+2 (d-5) \lambda\,L^2 \sigma \right)}
\nonumber\\
&&\times\Big[a_\star^{d+1}\, (d-1)\,(1 + 4\,\lambda)  \left(3 + 2\,\lambda\,L^2\,g_+^\star \right) 
\nonumber\\
&&\quad\; +\, 4 \,a_\star^{d-1}\,(d+1)\,(1 + 4\, \lambda )\, \lambda\, L^2\, \sigma
\nonumber\\
&&\quad +\, 4\, a_\star^{5-d} \left(5\,d-13 + 2 (d-3)\,\lambda\, L^2\, g_+^\star\right)\lambda\, L^2\, \mathcal{Q}^2 
\nonumber\\
&&\quad\; +\, 16\, a_\star^{3-d}\, (2 d-7)\, \lambda^2\, L^4\, \mathcal{Q}^2 \sigma \Big]\,.
\label{M_-}
\end{eqnarray}
Here we used $g_\pm^\star \equiv g_\pm (a_\star)$\,.
Then, we will find the solution of the functions $g_\pm^\star = g_\pm(a_\star)$ in terms of $a_\star$, $\lambda$, $L$, $d$ and $\mathcal{Q}$ via the $\Upsilon[g_\pm]$ functions. One finds
\begin{eqnarray}
\Upsilon[g_+^\star] &=& -\frac{1}{L^2} + g_+^\star + \lambda\,L^2\,(g_+^\star)^2 = \frac{\mathcal{M}_+^\star}{a_\star^{d-1}} -\frac{\mathcal{Q}^2}{a_\star^{2\,(d-2)}}
\nonumber\\
&\Rightarrow& -\frac{1}{L^2} + g_+^\star + \lambda\,L^2\,(g_+^\star)^2 = \mathcal{C}_1\,g_-^\star + \mathcal{C}_2\,,
\end{eqnarray}
and
\begin{eqnarray}
\Upsilon[g_-^\star] &=& -\frac{1}{L^2} + g_-^\star + \lambda\, L^2\,(g_-^\star)^2 = \frac{\mathcal{M}_-^\star}{a_\star^{d-1}} -\frac{\mathcal{Q}^2}{a_\star^{2\,(d-2)}}
\nonumber\\
&\Rightarrow& -\frac{1}{L^2} + g_-^\star + \lambda\,L^2\,(g_-^\star)^2 = \mathcal{C}_1\,g_+^\star + \mathcal{C}_2 \,,
\end{eqnarray}
where the coefficients $\mathcal{C}_{1,2}$ are given by
\begin{eqnarray}
\mathcal{C}_1 &=& \frac{4\, a_\star^{2(3-d)}(d-3)\,\lambda\, L^2\, \mathcal{Q}^2+ a_\star^2\, (d-1)\, (1+ 4\,\lambda)}{2\, a_\star^2 \,(d-1) + 4\, (d-5)\, \lambda\, L^2\, \sigma }
\\
\mathcal{C}_2 &=& \frac{(1 + 4\,\lambda)\left(3 \,a_\star^2\, (d-1) + 4\,(d+1)\,\lambda\, L^2\, \sigma \right)+ 8\,a_\star^{2(2-d)}\,(d-3)\,\lambda\, L^2\,\mathcal{Q}^2 \left(2\,a_\star^2+ 3\,\lambda\, L^2\, \sigma \right)}{4\,\lambda\, L^2 \left(a_\star^2 (d-1)+ 2\, (d-5)\, \lambda\, L^2\, \sigma \right)}
\nonumber\\
\end{eqnarray}
Solving above two equations simultaneously, we obtain the solutions of $g_\pm^\star$ as
\begin{eqnarray}
g_+^\star &=& -\,\frac{(1 + \mathcal{C}_1) + \sqrt{1 + 4\,\lambda - 2\,\mathcal{C}_1 - 3\,\mathcal{C}_1^2 + 4\,\mathcal{C}_2\,\lambda\, L^2}}{2\,\lambda\,L^2}
\\
g_-^\star &=& -\,\frac{(1 + \mathcal{C}_1) - \sqrt{1 + 4\,\lambda - 2\,\mathcal{C}_1 - 3\,\mathcal{C}_1^2 + 4\,\mathcal{C}_2\,\lambda\, L^2}}{2 \lambda\,L^4}\,.
\end{eqnarray}
We note that $g_-^\star$ has a good behavior (stable) for $\lambda\rightarrow 0$ while $g_+^\star$ gives infinite value (unstable) for $\lambda \rightarrow 0$\,. In addition, we need to study the phase transition between two manifolds of the spacetime, i.e., AdS (outer, $+$) to dS (inner, $-$) then the condition $g_+^\star \neq g_-^\star$ is necessary.
\begin{figure}[h]	
	\includegraphics[width=15cm]{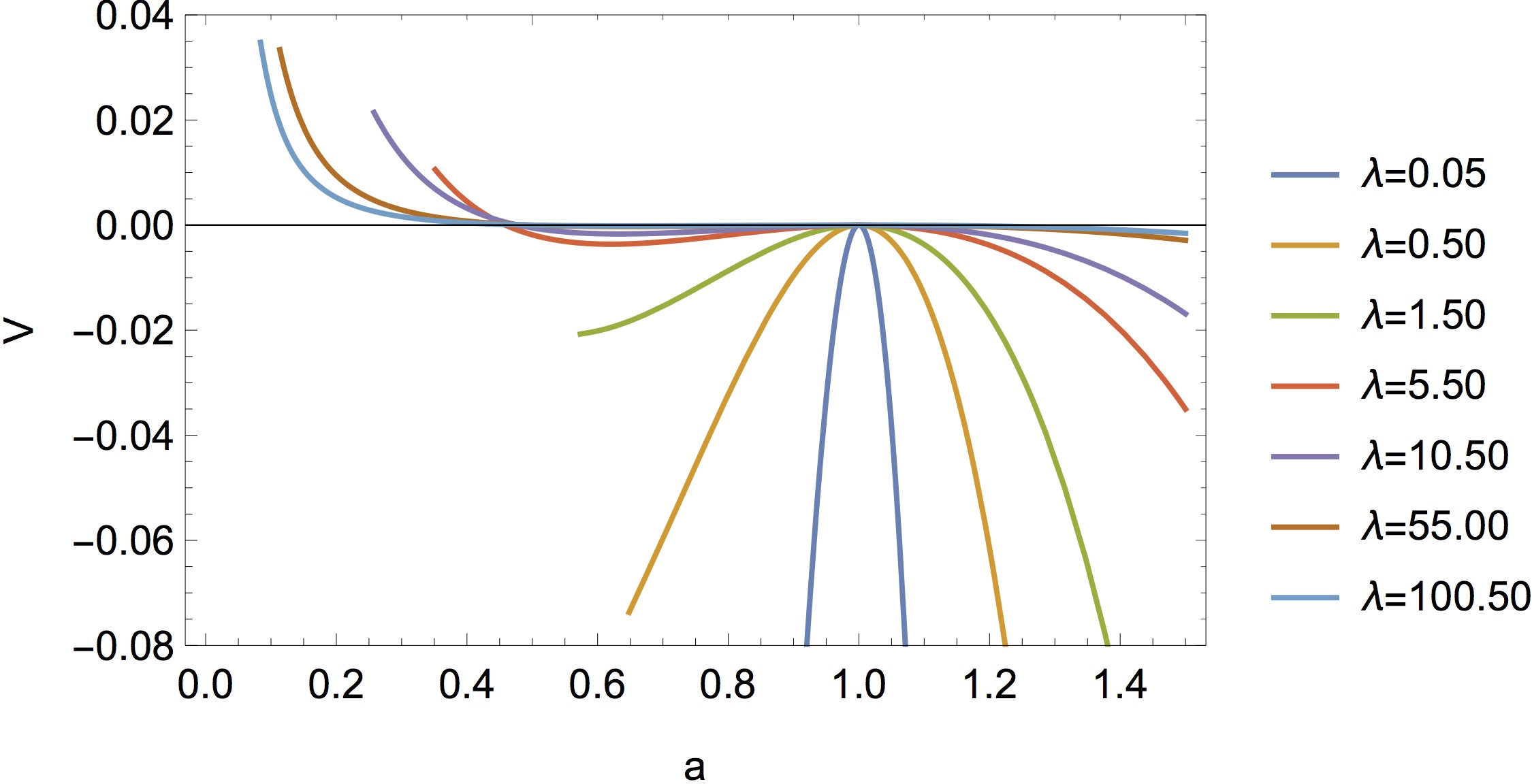}
	\centering
	\caption{The figure displays the shapes of the effective potential of thermalon in various values of the coupling $\lambda$ with $a_\star=1$, $L=1$, $d=5$, $\sigma=1$ and $\mathcal{Q}=0.5$. We found that there is no local minima of the effective potential at the thermalon position ($a_\star=1$) for any positive values of the $\lambda$ coupling. The thermalon always is unstable and gives the phase transition from AdS to dS spacetimes. }
	\label{thermalon-stability}
\end{figure}

To study the stability of the thermalon, we consider the bubble dynamics at the thermalon solutions at $a= a_\star$ and expand the junction condition in Eq. (\ref{junction-EMGB}) up to the first order as,
\begin{eqnarray}
\widetilde\Pi \approx \widetilde\Pi_\star  + \frac{\partial \widetilde\Pi_\star}{\partial H}\,\frac{\dot a^2}{a^2_\star} + \frac{\partial \widetilde\Pi_\star}{\partial a}\,(a-a_\star) +\frac12\,\frac{\partial^2 \widetilde\Pi_\star}{\partial a^2}\,(a - a_\star)^2 + \cdots\,,
\end{eqnarray}
at $a=a_\star$, one finds 
\begin{eqnarray}
\widetilde\Pi_\star = 0 = \frac{\partial \widetilde\Pi_\star}{\partial a}\,.
\end{eqnarray}
The junction condition above can be re-written as
\begin{eqnarray}
\frac12\,\dot a^2 + V_{\rm eff}(a) = 0\,,
\end{eqnarray}
where
\begin{eqnarray}
V_{\rm eff}(a) = \frac12\,k\,(a-a_\star)^2\,,\qquad k = \frac12\,a_\star^2\left( \frac{\partial \widetilde\Pi_\star}{\partial H}\right)^{-1}
\frac{\partial^2 \widetilde\Pi_\star}{\partial a^2}\,.
\label{V-eff-k}
\end{eqnarray}
It has been shown in Refs. \cite{Camanho:2015ysa,Hennigar:2015mco} that the sign of $k$ variable demonstrates the stability of the thermalon at $a=a_\star$. The thermalon configuration will be stable if $k$ greater than zero. On the other hand, $k<0$ gives the thermalon is unstable. Having used of the formula in Eq. (\ref{V-eff-k}), we need the thermalon to expand and then giving the phase transition of the bulk spacetime. This means that the $k<0$\,. With $|\mathcal{Q}|< |\mathcal{Q}_c|$ and $|\mathcal{Q}| = |\mathcal{Q}_c|$ limits, we have checked numerically and we found $k<0$ for all $\lambda >0$. One may conclude, in this case, that the effective charge $\mathcal{Q}$ does not change the stability of the thermalon configuration. This can be depicted by the shapes of the potential in the various positive values of the coupling, $\lambda > 0$ in figure \ref{thermalon-stability}. We find that there is no local minima at the thermalon configuration at $a_\star = 1$. The thermalon position locates on the top of the potential and it is unstable. At this point, the thermalon expands and then reaches the asymptotic region in a finite time and therefore changes the AdS to dS geometries of the whole spacetime. 

Furthermore, we consider the expansion of the bubble thermalon escaping to infinity. Considering the matching condition $\widetilde\Pi$ and keeping the first order of the $1/H$ expansion, we find (see \cite{Camanho:2013uda} for detail derivation in the the neutral case, $\mathcal{Q}=0$)
\begin{eqnarray}
H\approx \frac{1}{2(\mathcal{M}_+ - \mathcal{M}_-)}\left[ a^{d-1}\int_{g_-}^{g_+}\Upsilon[x]\,dx -\left( g_+\left[ \mathcal{M}_+ -\frac{\mathcal{Q}^2}{a^{d-3}}\right] - g_-\left[ \mathcal{M}_- -\frac{\mathcal{Q}^2}{a^{d-3}}\right]\right)\right].
\nonumber\\
\end{eqnarray}
At $a \to \infty$ limit, this gives $H\to \infty$. In addition,  we observe that $H\to \infty$ is a bit slower than the neutral case. 

\section{Gravitational phase transition}
\label{s3}
\subsection{Thermalon configurations, horizons and Nariai bound}
We come to the crucial part of this work. Before we move forward to the gravitational phase transition with the relevant thermodynamics quantities. The thermalon (bubble) location, $a_\star$, needs to ensure that it lies between the black hole radius, $(r_{B})$ inside the bubble and the cosmological horizon $(r_C)$. One can solve for $f(r_{H}) = 0$ in the function of $M_-^\star$ as,
\begin{eqnarray}
f_-(r_{H}) = 0\,,\;\Rightarrow \; g_-(r_{H}) = \frac{\sigma}{r_{BH}^2} \,,
\end{eqnarray}
where $r_H$ is the radius of the existent horizons of the spacetime.
The above equation gives
\begin{eqnarray}
\Upsilon_-\left[ \frac{\sigma}{r_{H}^2}\right] = \frac{\mathcal{M}_-^\star}{r_{H}^{d-1}} - \frac{\mathcal{Q}^2}{r_{H}^{2(d-2)}}\,,
\end{eqnarray}
where the expression of the $\mathcal{M}_-^\star \equiv \mathcal{M}_-^\star(g_\pm^\star,\,a_\star,\,\lambda,\,L^2,\,\mathcal{Q}^2)$ is given by Eq.(\ref{M_-}). More importantly, we will focus our study of the AdS to dS gravitational transition in $d=5$ and $\sigma=1$ (spherical geometry). Setting $f_-(r_H)=0$, the (de-Sitter branch, inner spacetime) horizons, $r_H$ can be obtained from the following equation 
\begin{eqnarray}
r_H^6 - L^2\,r_H^4 + L^2\,\big( \mathcal{M}_-^\star - \lambda\,L^2\big)\,r_H^2 - L^2\,\mathcal{Q}^2 = 0\,.
\label{cubic}
\end{eqnarray}
\begin{figure}[h]	
	\includegraphics[width=10cm]{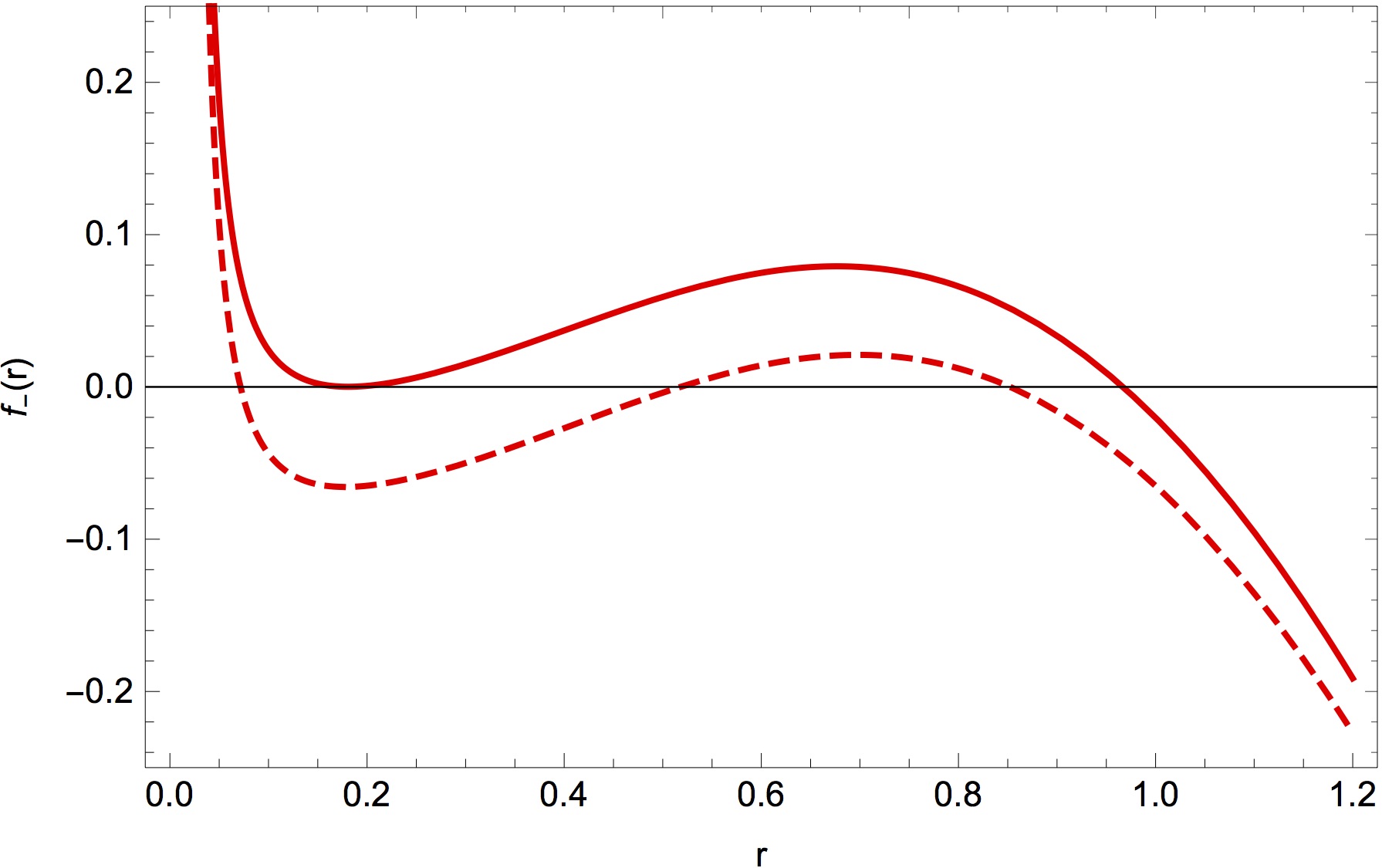}
	\centering
	\caption{The figure shows the existences of the event horizon and cosmological horizon for non-extremal ($|\mathcal{Q}|<|\mathcal{Q}_c|$) and extremal ($|\mathcal{Q}|=|\mathcal{Q}_c|$) cases. The dashed and solid lines represent three (inner, outer and cosmological horizons) and two (degenerated and cosmological horizons) horizons for non-extremal and extremal metric in Eq.(\ref{metric-f-pm}) respectively with $\lambda = 1$, $M=1.2$, $L=1$ and $\mathcal{Q}=0.001$.}
	\label{horizons}
\end{figure}
Having re-scaled $r_H^2\to u$\,, we find that the above equation reduces to a cubic equation of a variable $u$\,. In order to obtain three real solutions of the cubic equation, it requires the discriminant of the cubic equation less than zero. Using the standard technique, the solutions of Eq.(\ref{cubic}) for the horizons of the inner spacetime are given by,
\allowdisplaybreaks
\begin{eqnarray}
r_{H,1}^2 &=& -\frac{a}{3} + \frac{2\sqrt{-\,p}}{\sqrt{3}}\,\sin\left[ \frac13\,\arcsin\left( \frac{3\sqrt{3}\,q}{2\left(\sqrt{-\,p}\right)^3}\right)\right]\,,
\label{outer-horizon}\\
r_{H,2}^2 &=& -\frac{a}{3} - \frac{2\sqrt{-\,p}}{\sqrt{3}}\,\sin\left[ \frac13\,\arcsin\left( \frac{3\sqrt{3}\,q}{2\left(\sqrt{-\,p}\right)^3}\right) + \frac{\pi}{3}\right]\,,
\label{inner-horizon}\\
r_{H,3}^2 &=& -\frac{a}{3} + \frac{2\sqrt{-\,p}}{\sqrt{3}}\,\cos\left[ \frac13\,\arcsin\left( \frac{3\sqrt{3}\,q}{2\left(\sqrt{-\,p}\right)^3}\right) + \frac{\pi}{6}\right],
\label{cosmo-horizon}
\end{eqnarray}
where 
\begin{eqnarray}
q = \frac{2\,a^3}{3} -\frac{a\,b}{3} + c\,,\quad p= b -\frac{a^2}{3}\,,\quad  a = -\,L^2\,,\quad 
b = \left( \mathcal{M}_-^* -\lambda\,L^2 \right), \quad c = -\,L^2\,\mathcal{Q}^2\,. 
\end{eqnarray}
More importantly, the real positive values of $r_H$ in Eqs.(\ref{outer-horizon},\ref{inner-horizon},\ref{cosmo-horizon}) must satisfy the following conditions,
\begin{eqnarray}
\Delta = \frac{q^2}{4} + \frac{p^3}{27} < 0 \,.
\label{exist-horizon}
\end{eqnarray}
In addition, a critical charge, $\mathcal{Q}_c$ is determined by setting $\Delta = 0$ and it reads,
\begin{eqnarray}
\big| \mathcal{Q}_c\big| &=& \frac{L^2}{3\sqrt{3}}\left( -2 + \frac{9}{L^2}\big( \mathcal{M}_-^\star - \lambda\,L^2\big) + 2\left[ 1-\frac{3}{L^2}\big( \mathcal{M}_-^\star - \lambda\,L^2\big)\right]^{\frac32}\right)^{\frac12}\,,
\label{critical-Q}
\end{eqnarray}
with
\begin{eqnarray}
\lambda \geq \frac{\mathcal{M}_-^\star}{L^2} - \frac13\,.
\end{eqnarray}
To confirm the condition of the horizons, we also did numerical demonstration of the existences of the horizon in figure \ref{horizons} for non-extremal and extremal cases. It is well known for the charged solutions of the static spherical symmetry ($\sigma=1$) at the event horizon of the black hole ($r_B$) that if $\big| \mathcal{Q}_c\big| < \mathcal{Q}$ there are two horizon covering the singularity and the extremal black hole has single event horizon for $\big| \mathcal{Q}_c\big| = \mathcal{Q}$. On the other hand, if $\big| \mathcal{Q}_c\big| > \mathcal{Q}$ the black hole reveals the naked singularity. 

\begin{figure}[h]	
	\includegraphics[width=12.5cm]{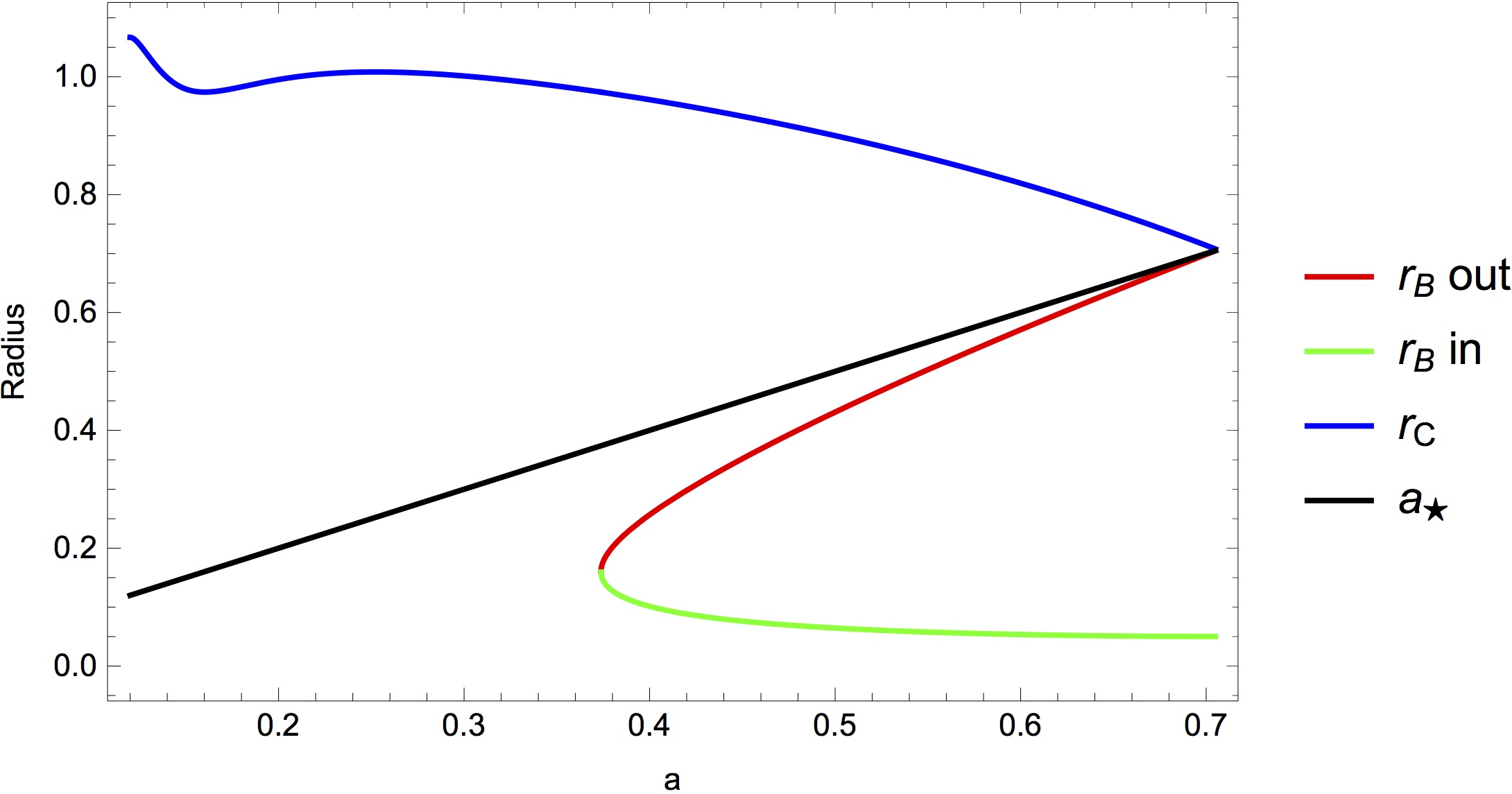}
	\centering
	\caption{A plot of an outer $r_{B}$ (red), an inner $r_{B}$ (green), $r_{C}$ (blue) and $a_{\star}$ (black) as functions of $a_{\star}$ for $\lambda = 0.20$, $L=1$, $d=5, \sigma=1$ and ${\cal Q}=0.025$. We observe that the bubble location, $a_{\star}$, is always found between the event horizons, $r_{B,{\rm in}}$ and $ r_{B,{\rm out}}$, and the cosmological horizon, $r_{C}$, until the outer event and cosmological horizons meet at the point where $a_{\rm Nariai}^{(\mathcal{Q})} = 0.7062$ given by Eq.(\ref{Nariai-Q}). This point is called the Nariai bound. For the neutral model, it is given by $a_{\rm Nariai}=\sqrt{3/\Lambda}=L/\sqrt{2}$ = $0.7074$ with $L=1$ \cite{Cai:2003gr,Camanho:2011rj}.} 
	\label{radius}
\end{figure}

Next we consider the smallest radius of the (inner) de-Sitter EMGB black hole, $r_S$. According to Eq.(\ref{metric-f-pm}), the smallest radius is the solution of the equation
\begin{eqnarray}
\big( 1 + 4\,\lambda\big)\,r_S^6 + 4\,\lambda\,L^2\,\mathcal{M}_-^\star\,r_S^4 - 4\,\lambda\,L^2\,\mathcal{Q}^2 =0\,
\end{eqnarray}
The solution of the $r_S$ corresponds to the curvature singularity or Cauchy horizon. Since the appearance of charge $\mathcal{Q}$ in the solutions makes the study of horizons more complicate. However, it has been shown in Ref. \cite{Thibeault:2005ha} that for the cosmological horizon ($r_C$) exists in the de-Sitter spacetime ($\Lambda \equiv 6/L^2 > 0$) and the Cauchy horizon is covered by the event horizon ($r_S < r_B$) with the following range of the parameters
\begin{eqnarray}
\frac{\mathcal{M}_-^\star}{L^2} - \frac13 < \lambda < \frac{\mathcal{M}_-^\star}{L^2}\,.
\label{r_H-cond}
\end{eqnarray}  
More importantly, there are the existences of the horizons in the EMGB gravity with de-Sitter spacetime. These has been proved in Ref.\cite{Torii:2005nh} that for $d=5$, $\sigma =1$ and $\Lambda \equiv 6/L^2 > 0$, there are three types of the horizons as inner, black hole and cosmological horizons. Types of horizons depend on the ranges of the mass and charge parameters \cite{Thibeault:2005ha,Torii:2005nh,Mishra:2019ged}. 
In the present work, we limit our study for the non-extremal black hole case due to the complicate relation between of the mass function $\mathcal{M}_-^\star$ and the charge, $\mathcal{Q}$ in Eq.(\ref{M_-}). Therefore, one can identify the radius of the outer, inner event and cosmological horizons from the solutions in Eqs.(\ref{outer-horizon},\ref{inner-horizon},\ref{cosmo-horizon}) as,
\begin{eqnarray}
r_{B,{\rm out}}^2 = r_{H,1}^2 \,,\qquad r_{B,{\rm in}}^2 = r_{H,2}^2 \,,\qquad r_C^2 = r_{H,3}^2 \,.
\end{eqnarray}
The expressions of above equations speculate that the thermalon position always lies between the event horizon and the cosmological horizon, $r_{B,{\rm in}} < r_{B,{\rm out}} < a_\star < r_C$\,. To demonstrate above speculation, we therefore plot all horizons and the thermalon configuration shown in figure \ref{radius} as a function of thermalon radius with $|\mathcal{Q}| < |\mathcal{Q}_c|$\,. One can see clearly that the outer (red line) and inner (green line) charged black hole event horizons are jointed smoothly and always covered by the thermalon radius (black line) while the cosmological horizon (blue line) is the largest radius and covers all horizons. The plot results in figure \ref{radius} are reproduced as Refs. \cite{Camanho:2015zqa,Camanho:2013uda,Hennigar:2015mco} when $\mathcal{Q}=0$ is taken into account. For EMGB gravity with the positive (bare) cosmological constant, in addition, our results are confirmed by Ref.\cite{Guo:2018exx} and it has been shown numerically that the event horizon is always covered by the cosmological horizon. Moreover, all ranges of the relevant parameters in the plots are numerically checked and they are obeyed the condition in Eq.(\ref{r_H-cond}). It is interesting to see in the case that the outer event horizon becomes larger until reaching the thermalon configuration and the cosmological horizon at some point. This point is called the Nariai bound and it is given by $a_{\rm Nariai} = \sqrt{3/\Lambda}=L/\sqrt{2}$ \cite{Camanho:2011rj,Cai:2003gr} for the neutral case. Interestingly, the Nariai bound for the charge case is given by
\begin{eqnarray}
a_{\rm Nariai}^{(\mathcal{Q})} = \sqrt{\frac{L^6 + L^2\,\mathcal{N} + \mathcal{N}^2}{6\,\mathcal{N}}}\,,\,\, \mathcal{N} = \left(L^6-54 L^2 Q^2 + 6  \sqrt{3\,L^4 Q^2 \,(27 Q^2 - L^4)}\right)^{\frac13}.
\label{Nariai-Q}
\end{eqnarray}
According to our results displayed in figure \ref{radius}, we discovered that the Nariai bound of the charge case with a small charge is very close to the neutral one. We will see in the latter that the gravitational phase transition will take place and it is satisfied by the Nariai bound.
\subsection{Thermodynamics quantities and critical phenomena of AdS to dS phase transition}
Next we step further to quantify the relevant thermodynamics quantities for studying the phase transition. In order to investigate the thermal AdS to dS black hole phase transition as done in Refs.\cite{Camanho:2015zqa,Camanho:2013uda,Camanho:2012da,Hennigar:2015mco} for the neutral model, we shall take a short overview of the mechanisms of the gravitational phase transition in the literature. The initial thermal AdS (outer geometry) will decay and transit to the black hole inside the dS spacetime (inner geometry) via the thermalon mediation. After the thermalon or the bubble (thin-shell) in the Euclidean sector is formed and it will expand eventually reaching the cosmological horizon entirely. At the end, the boundary of a whole spacetime is changed from AdS to dS geometries, i.e., the cosmological constant changes from negative to positive values. Therefore, the observer inside the cosmological horizon can measure the thermodynamics quantities of the dS spacetime. One may conclude that the thermalon changes the solutions from one branch to another via the phase transtion. More importantly, it has been shown that an reversible process for AdS to dS phase transition does not occur see more detail discussions in Refs.  \cite{Camanho:2015zqa,Camanho:2013uda,Sierra-Garcia:2017rni,Hennigar:2015mco}. For example, a so-called reentrant phase transition process happening in the study of black hole thermodynamics \cite{Altamirano:2013ane,Frassino:2014pha} is not possible.
The main purpose of this work is to investigate the phase transition profile of this scenario by including the static charge. In the following, we recall the main ingredients for this task. It has been proven and demonstrated in Refs.\cite{Camanho:2013uda} (see \cite{Camanho:2015ysa} for detail derivation) that in the canonical ensemble including the bulk (both inner and outer manifolds) and the surface actions, the Euclidean action of the thermalon configuration ($\mathcal{I}_E$) is related to the inverse Hawking temperature ($\beta_+$), mass ($\mathcal{M}_+$) of the external observer in the asymptotic thermal AdS and the entropy of the dS black hole. It reduces to a simple and compact form as,
\begin{eqnarray}
\mathcal{I}_E = \beta_+\,\mathcal{M}_+ + S\,.
\end{eqnarray}
Having used the on-shell regularization method by subtracting the thermal AdS space (outer branch solution) contribution as argued in Refs. \cite{Camanho:2015zqa,Camanho:2013uda,Hennigar:2015mco} for the neutral model, this leads to the (Gibbs) free energy of the thermalon configuration. It reads \cite{Camanho:2015zqa,Camanho:2013uda},
\begin{eqnarray}
F = \mathcal{M}_+ + T_+\,S\,,
\label{free-energy}
\end{eqnarray}
where $T_+ = 1/\beta_+$ is the Hawking temperature. In the latter, the free energy of the thermalon is compared to the thermal AdS space where the thermal AdS space is set to zero ($F_{\rm AdS} = 0$) because it was considered to be the background subtraction \cite{Camanho:2015zqa,Camanho:2013uda,Hennigar:2015mco,Sierra-Garcia:2017rni}.  Before we go further to quantify the relevant thermodynamics variables, it is worth noting that there are former five free parameters in the theory of the neutral case, i.e., $\mathcal{M}_\pm$, $T_\pm$ and $a_\star$\,. 

By using four conditions, there are two equations $V(a_\star)=0=V'(a_\star)$ from the configurations of the thermalon, Hawking temperature condition to avoid canonical singularity at the horizon, $T = f'(r_B)/4\,\pi$ and the matching temperature of the thermal circle at the thermalon configuration $\beta_+\sqrt{f_+(a_\star)}=\beta_-\sqrt{f_-(a_\star)}$. We found there is only free parameters and choose $T_+=1/\beta_+$\,. But, the inclusion of the vacuum static charge in this work gives an additional free parameter $\mathcal{Q}$\,. The Hawking temperature, $T_+$ is give by
\begin{eqnarray}
T_+ = \sqrt{\frac{f_+(a_\star)}{f_-(a_\star)}}\,T_-\,,
\label{T_+}
\end{eqnarray}
where the $T_-$ is the Hawking temperature of the inner dS black hole in EMGB gravity and it is determined with $d$-dimension and general spatial curvature $\sigma$ by \cite{Cai:2003kt},  
\begin{eqnarray}
T_-  =  \Bigg[\sum_{k=0}^{2}(d-1 -2k)\,\sigma\,c_k \left(\frac{\sigma}{r_B^2}\right)^{k-1} - (d-3)\frac{\mathcal{Q}_c^2}{r_B^{2d-6}}\Bigg]\Bigg[4\,\pi\, r_B\sum_{k=0}^{2}k\,c_k\left(\frac{\sigma}{r_B^2}\right)^{k-1}\Bigg]^{-1}.
\label{T_-}
\end{eqnarray}
The entropy $S$ is given by \cite{Cai:2003kt},
\begin{eqnarray}
S = 4\,\pi \sum_{k=0}^{2}\frac{k\,c_k}{d-2\,k}\left( \frac{\sigma}{r_B^2}\right)^{k-1}\,.
\label{entropy}
\end{eqnarray}
We note that the entropy of the charged black hole has the same form as the neutral black hole \cite{Camanho:2015zqa,Camanho:2013uda,Hennigar:2015mco}. In addition, the mass parameter $\mathcal{M}_-^\star$ is given by Eq.(\ref{M_+}). Having used the outer event horizon in Eq.(\ref{outer-horizon}) and substituted into Eqs.(\ref{M_+},\ref{T_+},\ref{entropy}), we obtain all building blocks of the thermodynamics quantities as function of thermalon radius and we are ready to study the thermalon properties and the gravitational phase transitions in the thermodynamics phase space.  
\begin{figure}[h]	
	\includegraphics[width=10cm]{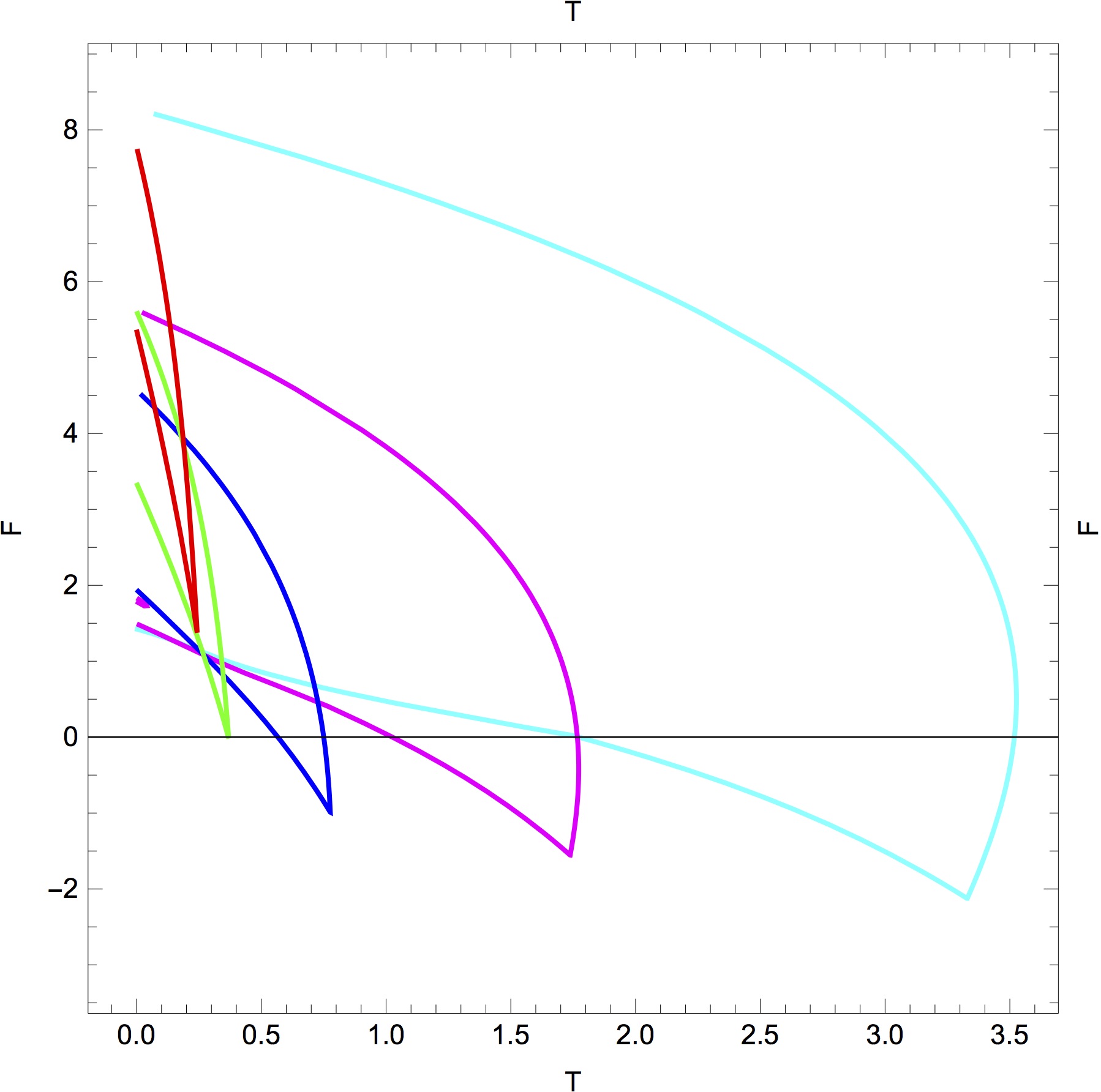}
	\centering
	\caption{The figure displays free energy $F$ of the thermalon configuration as a function of the temperature $T=\beta^{-1}_{+}$ for several values of the coupling $\lambda$. We have used $L=1$, $\sigma=1$, $d=5$, and ${\cal Q}=0.175$. From right to left: $\lambda=0.05$ (cyan), $\lambda=0.10$ (pink), $\lambda=0.25$ (blue), $\lambda=0.65$ (green) and $\lambda=1.20$ (red). For each value of the coupling $\lambda$, the upper branch beyond the cusp is unphysical where it corresponds to $\Pi^+ = -\,\Pi^-$ solutions while the lower branch is the physical solutions of $\Pi^+=\Pi^-$.}
	\label{FT}
\end{figure}

The free energy in Eq.(\ref{free-energy}) plays the crucial role for investigating the phase transition. The behavior of the free energy is very interesting and is influenced by both the coupling $\lambda$ and the charge ${\cal Q}$. Note that a cusp structure for the given values of the coupling $\lambda$ and the charge $\mathcal{Q}$ indicates the lowest value of the free energy $F$ which is lower than the thermal AdS space ($F_{\rm AdS} = 0$, in this work it is zero as mentioned earlier) at the same temperature. Then the thermolon will jump to the dS branch solution and changes the boundary from AdS to dS asymptotics resulting the discontinuity (cusp) of the free energy $F$ at the maximum temperature of the physical branch solutions. This leads to the zeroth-order phase transition see \cite{Camanho:2015zqa,Sierra-Garcia:2017rni} for detailed discussion. To investigate the AdS to dS phase transition, we consider figure \ref{FT} displaying free energy $F$ of the thermalon configuration as a function of the temperature $T=\beta^{-1}_{+}$ for several values of the coupling $\lambda$ with the fixed value of the charge $\mathcal{Q}$. We have used $L=1$, $\sigma=1$, $d=5$, ${\cal Q}=0.15$. From right to left: $\lambda=0.05$ (cyan), $\lambda=0.10$ (pink), $\lambda=0.25$ (blue), $\lambda=0.65$ (green) and $\lambda=1.20$ (red). More importantly, for each value of the $\lambda$ coupling of the $T$ vs $F$ phase diagram in figure \ref{FT}, we point out that the upper branch beyond the cusp is unphysical branch solutions where it corresponds to $\Pi^+ = -\,\Pi^-$ solutions of the $V(a_\star)=0=V'(a_\star)$ conditions while the lower branch is the physical solutions $\Pi^+ = \Pi^-$, see more discussions in Ref.\cite{Hennigar:2015mco}. We notice that for various ranges of temperatures the free energy at the maximum temperature of the (physical) branch is negative (i.e., less than free energy of the thermal AdS, $F_{\rm AdS}=0$) implying the possibility of thermalon mediated phase transition \cite{Camanho:2015zqa,Sierra-Garcia:2017rni,Hennigar:2015mco}.  Note that interpolating the cusp structures of the free energy at the maximum temperature of the physical solution corresponds to the curve of the Nariai bound of the dS branch solution \cite{Hennigar:2015mco}. Additionally, we observe that the range of temperatures over which these transitions emerge increases as the coupling $\lambda$ is given smaller with a small charge required. Moreover, thermalon mediated phase transitions are possible over a wide range of temperature for smaller values of the coupling $\lambda$ and the condition in Eq.(\ref{r_H-cond}) is still valid. However, for the given charge value $\mathcal{Q}=0.15$ in figure \ref{FT}, the phase transition is not possible for the coupling $\lambda \gtrsim 0.65$ see green and red lines where the cusp structures of the free energy occur for $F \geq 0$.
\begin{figure}[h]	
	\includegraphics[width=10cm]{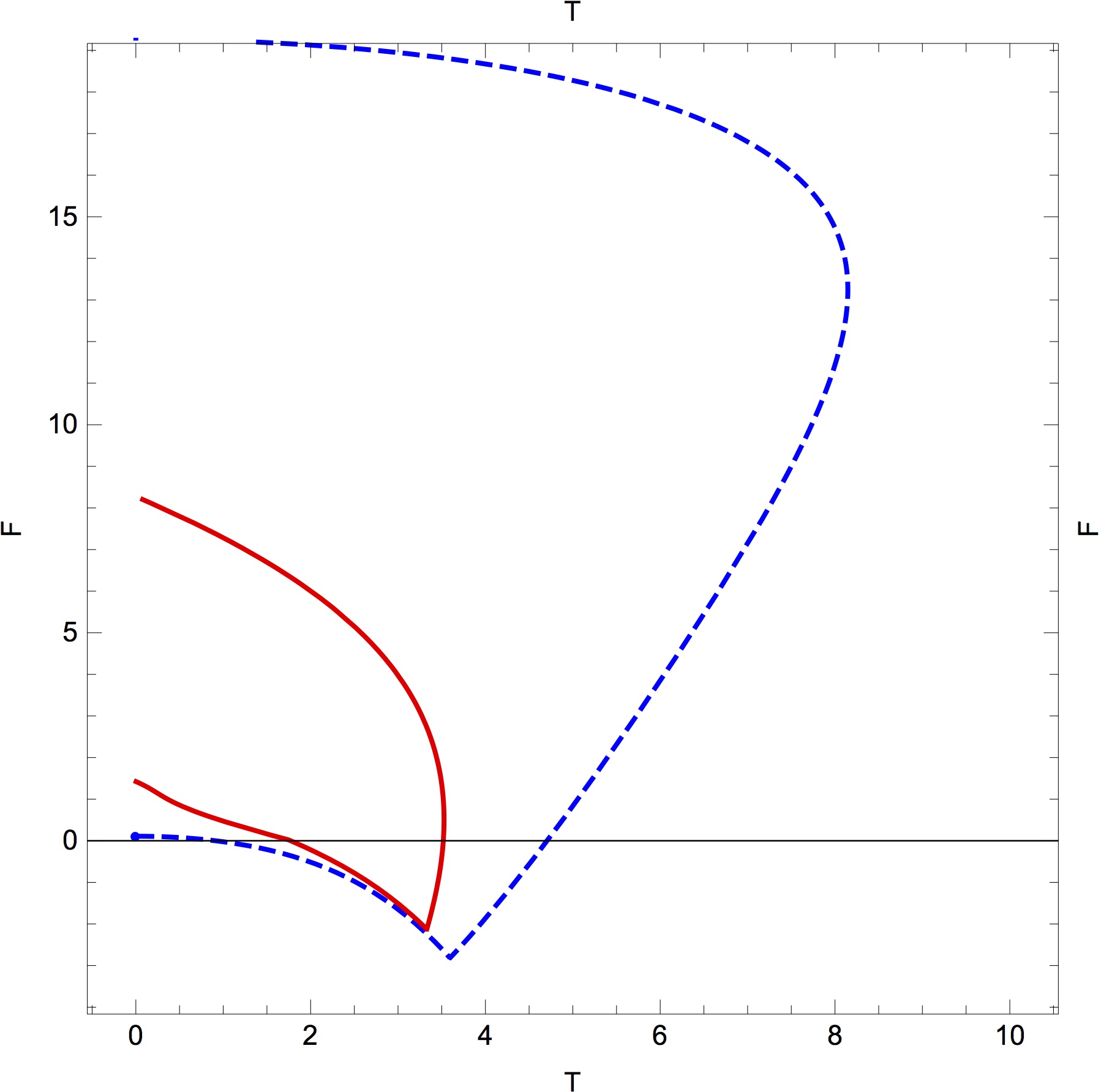}
	\centering
	\caption{The figure displays free energy $F$ of the thermalon configuration as a function of the temperature $T=\beta^{-1}_{+}$. We have used $L=1$, $\sigma=1$, $d=5$ and $\lambda = 0.05$. The red line shows $F$ vs $T$ of a charged case with ${\cal Q}=0.15$, while the dashed blue line indicates $F$ vs $T$ of a neutral case with ${\cal Q}=0$.}
	\label{FTqnoq}
\end{figure}
In contrast of the study of the AdS to dS phase transition in the neutral case, the phase transition takes the place for the critical value of the coupling with $\lambda = 1.138$ \cite{Camanho:2015zqa}. Inclusion of the charge, however, we find that there is no phase transition (i.e., the free energy is greater than zero) for the critical value of the coupling $\lambda = 1.138$. We then extensively study by comparing the plot of the free energy $F$ of the thermalon configuration vs the temperature $T=\beta^{-1}_{+}$ between the charge and the neutral models with $L=1$, $\sigma=1$ and $d=5$. The red line shows $F$ vs $T$ of a charged case with $\lambda=0.05$ and ${\cal Q}=0.15$, while the dashed blue line indicates $F$ vs $T$ of a neutral case with $\lambda=0.05$ and ${\cal Q}=0$. It is worth noting from figure \ref{FTqnoq} that the phase transitions of the charged case, ${\cal Q}\neq 0$, is possible in which the required maximum temperature of the physical branch is lower than that of the neutral case, ${\cal Q}=0$. Interestingly, we notice that the critical (maximum) temperature and coupling $\lambda$ of the phase transitions are modified when adding the charge. We have also checked that at a fixed value of $\lambda$ the critical temperature of the phase transition decreases when the charge gradually increases followed by a condition $|{\cal Q}|<|{\cal Q}_{c}|$. This phenomena is similar to the physical situation in the condensed matter physics, i.e., adding a charge as an impurity substitution. For instance, the conventional superconductivity is a single normal impurity with a small concentrations. Increasing the size of the impurity in a fixed-size host superconductor gives decreasing critical temperature of the host superconductor \cite{Ghosal:1998,Xiang:1995}. 


\section{Conclusion}

In this work, we have revisited the toy model of the AdS to dS phase transition in higher-order gravity proposed by Ref.\cite{Camanho:2015zqa}. Notice that the gravitational phase transition for the neutral case in the vacuum solutions has been extensively studied in the literature. It was proposed that the thermalon, the Euclidean spherical thin-shell, plays crucial role of the phase transition as already mentioned in the section \ref{s3}. In the other words, the thermalon changes the branches of the solutions from one branch to another via the thermal phase transition. This phenomenon is a generalization of the Hawking-Page phase transition and it is expected to be a generic behavior of the phase transition in higher-order gravity. We then extend the study of the AdS to dS phase transition by adding the Maxwell field as an impurity substitution for investigating the profile of the phase transition in this framework. We therefore focus on EMGB gravity in this work. The junction condition in the EMGB theory is also constructed and this leads to the effective potential of the thermalon in the non-extremal case ($|\mathcal{Q}| < |\mathcal{Q}_c|$). We found that the inclusion of the Maxwell field (the static charge) does not change the dynamics and stability of the themalon as shown in the section \ref{s2} except the existences of the effective potential. As expected, we found that there are three horizons of the interior space existing in this scenario, i.e., outer event, inner event and cosmological horizons. The thermalon radius always located between outer event and cosmological horizons.

In addition to the study of phase transition in the thermodynamic phase space, the behaviors of the (Gibbs) free energy ($F$) vs temperature ($T_+$) exhibit a possibility of the phase transition with the presence of the static charge. The phase transition takes place when the free energy is lower than the thermal AdS space ($F_{\rm AdS}=0$) at the maximum temperature of the (physical) branch solutions. This leads to the thermalon transition from the AdS to dS branch solutions. Comparing to the neutral case, we found that the inclusion of the static charge affects the critical higher-order coupling $\lambda$ and the maximum value of the temperature of the phase transitions. For a fixed value of the charge $|\mathcal{Q}|<|\mathcal{Q}_c|$, the critical (maximum) temperature and the coupling $\lambda$ of the thermalon transition are lower than the neutral case. When fixing a value of $\lambda$, the maximum temperature of the physical branch decreases if the static charge increases. According to the results present in this work, we conclude that the inclusion of the Maxwell field (static charge) in the gravitational phase transition behaves in the same way as that of the impurity substitution in condensed matter physics as the zeroth order phase transition. Moreover, adding matter field in higher-order gravity does not change the profile of the phase transition. Our results agree with the claim that the gravitational AdS to dS phase transition is a generic transition mechanism of the theories of higher-order gravity.  

Based on our analysis, inclusion of more complex fields, e.g., adding matter fields, might gain a deeper understanding of the dS/CFT structure. Some existing fields in string theory might reveal rich phenomena and new features of the gravitational phase transition, for instance, the three form field is one of these interesting substitutions and it is worth further investigating. 

\acknowledgments
We are grateful to J. Anibal Sierra-Garcia for enlightening discussions and thorough suggestions of the detailed calculations. This work is financially supported by Walailak University under grant no.WU-CGS-62001.

\end{document}